# Topological impact of a simple self-replication geometric structure with great application potential in vacuum pumping and photovoltaic industry

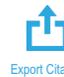

View Online    Export Citation    CrossMark

Xueli Luo[a]) and Christian Day

**AFFILIATIONS**

Karlsruhe Institute of Technology, Institute for Technical Physics, 76021 Karlsruhe, Germany

[a])Author to whom correspondence should be addressed: Xueli.Luo@kit.edu

**ABSTRACT**

Topological effects exist from a macroscopic system such as the universe to a microscopic system described by quantum mechanics. We show here that an interesting geometric structure can be created by the self-replication procedure of a square with an enclosed circle, in which the sum of the circles' area will remain the same but the sum of the circumference will increase. It is demonstrated by means of Monte Carlo simulations that these topological features have a great impact on the vacuum pumping probability and the photon absorption probability of the active surface. The results show significant improvement of the system performance and have application potential in vacuum pumping of large research facilities such as a nuclear fusion reactor, synchrotron, and in the photovoltaic industry.

## I. INTRODUCTION

Topological features are related to the geometric structure of a system.[1] Square and circle are two simple shapes in two-dimensional geometry. Suppose there is one square with the side length of a. First, one enclosed (inscribed) circle is put into the square and so the square is divided into four disconnected corners (spaces). In the second step, the original square is divided into four identical squares and each with an enclosed circle. In the next steps, this one-to-four self-replication procedure (one-to-two in each dimension) just repeats, as shown in Fig. 1.

This self-replication procedure is under the constraint of the original square. In each step N, the number of circles M, the curvature $\kappa$ of the circle, the number of the disconnected spaces $\chi$, the total area of the circles Ao, and the total circumferences of the circles L are as follows:

$$M = 4^{N-1}, \quad N = 1, 2, 3, \ldots, \quad \kappa = \frac{2^N}{a}, \quad N = 1, 2, 3, \ldots,$$
$$\chi = (2^{N-1} + 1)^2, \quad N = 1, 2, 3, \ldots, \quad (1)$$
$$Ao = \frac{1}{4}\pi a^2, L = 2^{N-1}\pi a, \quad (N = 1, 2, 3, \ldots).$$

These relations reveal two interesting topological features: (1) under the same constraint of the original two-dimensional square, the zero-dimensional and one-dimensional measurements, such as the number of circles M, the number of the disconnected spaces $\chi$, and the circumference L, could be unlimited; (2) the difference between bound and limit. Actually, the natural bound of the area here is $a^2$, but the total area of the circles is always unchanged as $Ao = \frac{1}{4}\pi a^2$ in each self-replication step N. A self-replication may not be confused with a fractal process, e.g., the Wallis sieve, where smaller and smaller structures are generated while the big original structure remains.

This self-replication procedure could be easily extended to the three-dimensional case, in which the original cube is filled with an enclosed (inscribed) sphere and is divided into eight identical smaller cubes filled with smaller enclosed spheres in the next step (also one-to-two in each dimension), and so on. The number of spheres M, the curvature $\kappa$ of the sphere, the number of the disconnected spaces $\chi$, the total volume of the spheres V, and the total surface areas of the spheres Ao are given as follows:

$$M = 8^{N-1}, \quad N = 1, 2, 3, \ldots, \quad \kappa = \frac{2^N}{a}, \quad N = 1, 2, 3, \ldots,$$
$$\chi = (2^{N-1} + 1)^3, \quad N = 1, 2, 3, \ldots, \quad V = \frac{1}{6}\pi a^3, \quad (2)$$
$$Ao = 2^{N-1}\pi a^2, \quad (N = 1, 2, 3, \ldots).$$

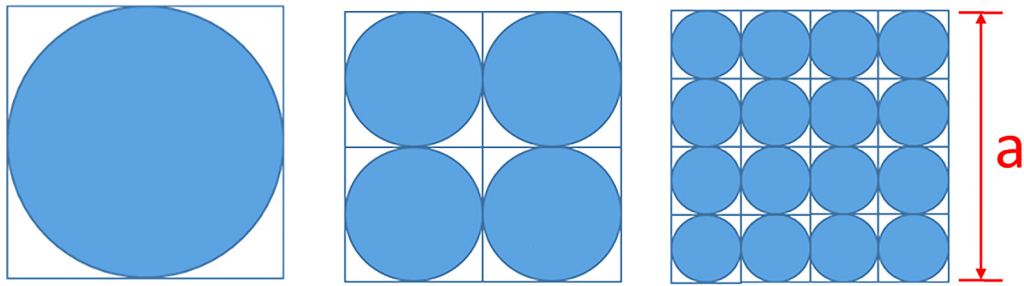

**FIG. 1.** Self-replication in the first steps from N = 1 to N = 3.

Astonishing enough is that the total volume of all spheres V is kept the same in each self-replication step as other measurements, such as the total surface areas of the spheres Ao, are increasing. However, note that in different self-replication steps, neither the circles in the two-dimensional case nor the spheres in the three-dimensional case are the same since the curvatures of them, which are equal to the reciprocal of the radius in these two cases, become greater as N increases, and, generally speaking, the curvature is an important physical parameter.[2]

In this paper, we will focus on the 2D self-replication procedure and investigate its impact on the system performance by systematic Monte Carlo simulations.

## II. PUMPING PROBABILITY

Usually, the requirement of the vacuum pumping is to provide high pumping speed for a given gas load under the system geometric constraint. The essential topological features of the aforementioned 2D self-replication procedure are that the total circumferences of the circles will increase as their total areas are kept unchanged under the constraint of the original square. This provides us an opportunity to exploit the third dimension perpendicular to the plane. In the first step N = 1, the original square is supposed to be an active pumping surface $A_s = a^2$ (a = 200 mm in simulation), and the enclosed circle is extended with a tube in the third dimension perpendicular to the plane as illustrated in Fig. 2. As given in Eq. (1), the number of tubes M in the system will increase dramatically as N increases.

The configuration shown in Fig. 2 is similar to the typical configuration to connect a pump to a vacuum system, and the pumping speed S of the system is

$$S = \frac{1}{4} A_s \langle v \rangle w, \qquad (3)$$

where $\langle v \rangle$ is the average velocity of the gas molecules in equilibrium, i.e., described by a Maxwell–Boltzmann distribution and w is the pumping probability of the system.[3] The particle flow rate corresponding to the known pumping speed S and particle density n is Q = n × S. When w = 1 in the limit case, $Q = \frac{1}{4} A_s \langle v \rangle n$ correlates with the particle flux onto the surface $A_s$ per unit time from one side. Note the difference with the typical configuration to connect a pump to a vacuum system, where the wall of the facility is usually made of steel and has no pumping effect (zero pumping probability), and $A_S$ in Eq. (1) will reduce to Ao, which is the inlet area of the pump.

In the free molecular flow regime, the pumping probability w of the system can be simulated by the test particle Monte Carlo (TPMC) simulation code such as MOLFLOW+.[4,5] In this study, we will use the code PROVAC3D developed by ourselves, which has been used in different research projects and successfully parallelized.[6–9] The length of the tube $T_L$ is assigned to be 0.25a, 0.5a, 0.75a, and a. Every tube includes the bottom. The sticking coefficient α of the original square, the tube wall, and the bottom is assigned to be 0.01, 0.02, and 0.03, respectively. The reflection from the surface is assumed to be a diffuse reflection. In order to distinguish the inlet from absorbing surfaces in the simulation, it is convenient to add a very short rectangular connection duct without pumping effect and with a length of 5‰ to the side length of the square is modeled as the reference case. In order to have precise simulation results, the simulations were carried out with $10^{12}$ or $10^{13}$ test particles on the supercomputer Marconi by using 640–16 000 cores in parallel. The most important trick is to completely avoid the pseudopenetration of the test particles between adjacent tubes at their tangent point as a result of the inevitable numerical errors in the simulation.

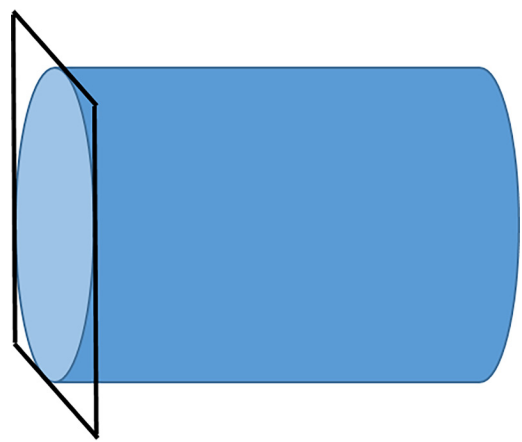

**FIG. 2.** Potential to exploit the dimension perpendicular to the plane in the first step N = 1.

Alternatively, the effective pumping probability w of this special system can be calculated by another method. When all tubes in each self-replication step N are independent and identical, the ratio of the total inlet area of all tubes to the area of the original square is always $\pi/4$, and w of the system for given $\alpha$ can be written as

$$w = k_1 w_{tube}(\alpha) + k_2 \alpha = \frac{1}{4}\pi w_{tube}(\alpha) + \left(1 - \frac{1}{4}\pi\right)\alpha, \quad (4)$$

with $w_{tube}(\alpha)$ being the pumping probability of each tube. In this way, the effect from the short duct will be automatically excluded and the simulation simplified since $w_{tube}(\alpha)$ is only dependent on the length to diameter ratio R in the free molecular flow regime. The weighing factors $k_1 = \frac{\pi}{4}$ and $k_2 = 1 - k_1 = 1 - \frac{\pi}{4}$ represent ratios of the total opening area for the bundle of parallel tubes and the unchanged flat area to the area $A_s$ of the original square, respectively.

The system has been simulated in both methods. Because they have very good agreement with each other, Table I only lists the simulation results in the first method that directly simulates the system of M tubes. In the table, the total area of the tube walls is $A = L \times T_L$, with L given by Eq. (1), R is the tube length to diameter ratio, and $\delta = (w - \alpha)/\alpha$ is the relative increase of w.

The notation of w in the table includes its significant digits and the uncertainty of the last digit in parenthesis. If total $N_{mc}$ test particles are simulated by TPMC in one case and $n_{ab}$ particles are absorbed by the active surfaces, then w is given by the ratio

$$w = \frac{n_{ab}}{N_{mc}}. \quad (5)$$

**TABLE I.** Simulation results of the pumping probability w and corresponding relative increase $\delta$.

|  | A | $\alpha = 0.01$ | $\delta$ | $\alpha = 0.02$ | $\delta$ | $\alpha = 0.03$ | $\delta$ |
|---|---|---|---|---|---|---|---|
| Reference case without tube | — | 0.009 999 5(1) | — | 0.019 997 9(1) | — | 0.0299 950(2) | — |
| N = 1, M = 1 |  |  |  |  |  |  |  |
| $T_L = 0.25a$, R = 0.25 | $0.25\pi a^2$ | 0.017 683 4(1) | 0.77 | 0.035 033 8(2) | 0.75 | 0.052 062 3(2) | 0.74 |
| $T_L = 0.5a$, R = 0.5 | $0.5\pi a^2$ | 0.025 224 6(2) | 1.52 | 0.049 523 0(2) | 1.48 | 0.072 950 4(3) | 1.43 |
| $T_L = 0.75a$, R = 0.75 | $0.75\pi a^2$ | 0.032 584 4(2) | 2.26 | 0.063 335 4(2) | 2.17 | 0.092 417 7(3) | 2.08 |
| $T_L = a$, R = 1 | $\pi a^2$ | 0.039 738 6(2) | 2.97 | 0.076 406 6(3) | 2.82 | 0.110 377 6(3) | 2.68 |
| N = 2, M = 4 |  |  | — |  | — |  | — |
| $T_L = 0.25a$, R = 0.5 | $0.5\pi a^2$ | 0.025 22 42(2) | 1.52 | 0.049 522 1(2) | 1.48 | 0.072 948 4(3) | 1.43 |
| $T_L = 0.5a$, R = 1 | $\pi a^2$ | 0.039 738 2(2) | 2.97 | 0.076 404 5(3) | 2.82 | 0.110 374 5(3) | 2.68 |
| $T_L = 0.75a$, R = 1.5 | $1.5\pi a^2$ | 0.053 353 5(2) | 4.34 | 0.100 179 5(3) | 4.01 | 0.141 708 2(3) | 3.72 |
| $T_L = a$, R = 2 | $2\pi a^2$ | 0.065 953 9(2) | 5.60 | 0.120 728 1(3) | 5.04 | 0.167 191 6(4) | 4.57 |
| N = 3, M = 16 |  |  | — |  | — |  | — |
| $T_L = 0.25a$, R = 1 | $\pi a^2$ | 0.039 737 7(2) | 2.97 | 0.076 402 9(3) | 2.82 | 0.110 369 6(3) | 2.68 |
| $T_L = 0.5a$, R = 2 | $2\pi a^2$ | 0.065 951 7(2) | 5.60 | 0.120 722 9(3) | 5.04 | 0.167 181 6(4) | 4.57 |
| $T_L = 0.75a$, R = 3 | $3\pi a^2$ | 0.087 882 5(3) | 7.79 | 0.152 688 0(4) | 6.63 | 0.203 274 0(4) | 5.78 |
| $T_L = a$, R = 4 | $4\pi a^2$ | 0.105 478 8(3) | 9.55 | 0.174 445 1(4) | 7.72 | 0.224 832 1(4) | 6.49 |
| N = 4, M = 64 |  |  | — |  | — |  | — |
| $T_L = 0.25a$, R = 2 | $2\pi a^2$ | 0.065 949 9(2) | 5.59 | 0.120 717 2(3) | 5.04 | 0.167 168 4(4) | 4.57 |
| $T_L = 0.5a$, R = 4 | $4\pi a^2$ | 0.105 473 8(3) | 9.55 | 0.174 431 6(4) | 7.72 | 0.224 810 2(4) | 6.49 |
| $T_L = 0.75a$, R = 6 | $6\pi a^2$ | 0.129 484 5(3) | 11.95 | 0.197 840 6(4) | 8.89 | 0.244 364 6(4) | 7.15 |
| $T_L = a$, R = 8 | $8\pi a^2$ | 0.142 835 3(3) | 13.28 | 0.207 273 0(4) | 9.36 | 0.250 665 4(4) | 7.36 |
| N = 5, M = 256 |  |  | — |  | — |  | — |
| $T_L = 0.25a$, R = 4 | $4\pi a^2$ | 0.105 467 7(3) | 9.55 | 0.174 415 5(4) | 7.72 | 0.224 783 6(4) | 6.49 |
| $T_L = 0.5a$, R = 8 | $8\pi a^2$ | 0.142 825 8(3) | 13.28 | 0.207 251 2(4) | 9.36 | 0.250 633 0(4) | 7.35 |
| $T_L = 0.75a$, R = 12 | $12\pi a^2$ | 0.153 597 6(4) | 14.36 | 0.212 447 5(4) | 9.62 | 0.253 354 4(4) | 7.45 |
| $T_L = a$, R = 16 | $16\pi a^2$ | 0.156 476 0(4) | 14.65 | 0.213 281 2(4) | 9.66 | 0.253 674 4(4) | 7.46 |
| N = 6, M = 1024 |  |  | — |  | — |  | — |
| $T_L = 0.25a$, R = 8 | $8\pi a^2$ | 0.142 814 6(3) | 13.28 | 0.207 227 6(4) | 9.36 | 0.250 597 6(4) | 7.35 |
| $T_L = 0.5a$, R = 16 | $16\pi a^2$ | 0.156 462 7(4) | 14.65 | 0.213 256 9(4) | 9.66 | 0.253 637 9(4) | 7.45 |
| $T_L = 0.75a$, R = 24 | $24\pi a^2$ | 0.157 442 4(4) | 14.74 | 0.213 429 4(4) | 9.67 | 0.253 691 0(4) | 7.46 |
| $T_L = a$, R = 32 | $32\pi a^2$ | 0.157 520 7(4) | 14.75 | 0.213 437 7(4) | 9.67 | 0.253 692 2(4) | 7.46 |
| N = 7, M = 4096 |  |  | — |  | — |  | — |
| $T_L = 0.25a$, R = 16 | $16\pi a^2$ | 0.156 451 2(4) | 14.65 | 0.213 234 1(4) | 9.66 | 0.253 605 4(4) | 7.45 |
| $T_L = 0.5a$, R = 32 | $32\pi a^2$ | 0.157 508 1(4) | 14.75 | 0.213 415 5(4) | 9.67 | 0.253 660 8(4) | 7.46 |
| $T_L = 0.75a$, R = 48 | $48\pi a^2$ | 0.157 516 4(4) | 14.75 | 0.213 416 1(4) | 9.67 | 0.253 660 3(4) | 7.46 |
| $T_L = a$, R = 64 | $64\pi a^2$ | 0.157 516 7(4) | 14.75 | 0.213 415 7(4) | 9.67 | 0.253 660 6(4) | 7.46 |

Because $N_{mc}$ is a fixed number and only $n_{ab}$ is of stochastic nature with statistical error, the variance (uncertainty) of w can be written as

$$\sigma(w) = \frac{\sigma(n_{ab})}{N_{mc}}, \quad (6)$$

and $\sigma(n_{ab})$ can be estimated by assuming a binominal distribution of the test particles in the simulation,[10]

$$\sigma(n_{ab}) \approx \sqrt{N_{mc} w (1-w)}. \quad (7)$$

We noticed that the pumping probabilities w of the reference cases without tube have very small deviations from the values of the initial sticking coefficients. This means that the conductance limitation of the short duct is negligible, and the improvement of the pumping probability comes from the self-replication procedure. In fact, the effect of the duct had also been checked if its length is 100 times shorter and the simulation results of w would only have a small relative change of a few percent in the positive direction. Moreover, it is found that for given sticking coefficient α, the increase of w is actually dependent on the active pumping area added in the system, which is the total area A of the tube walls. However, w will approach a maximum value as A increases. Because all tubes in each self-replication step N are identical in our simulation model, there exists a simple relationship between $A/A_s$ and the length to diameter ratio of the tube R, i.e., $A/A_s = \pi R$. So, w is actually determined by R, which has been proved by the simulation results in Table I.

Figure 3 shows (a) the relation of w to R and α and (b) its relative increase δ. In the figures, R = 0 corresponds to the original square with the pumping probability $w_0 = \alpha$. It can be clearly seen that the maximum value of w depends on the original α, and the greater the original α is, the greater the maximum value of w is. However, the greater the original α is, the faster the relative increase will approach its maximum. In our cases, if the original α = 0.01, the maximum value of w is 0.1575, with a relative increase about δ = 14.75 times; if the original α = 0.03, the maximum value of w is 0.2536, with a relative increase about δ = 7.46 times.

Please note that the side length of the original square could be arbitrary. This means that the pumping probability w of the system is determined by the ratio of the area of the tube walls to the area of the original square $A/A_s$. A larger original square is in favor of a higher pumping speed S and the tubes of larger diameters can be attached. However, the effective w of the system is only determined by the ratio $A/A_s$ (or R) after more and more active pumping surfaces are added, no matter how they would be arranged with fewer and longer tubes or with more and shorter tubes. This is reasonable if a smaller square is considered as one part of a larger original square. Moreover, this relationship would be the same if the tubes would be added with a different process other than the self-replication procedure in our study, for example, with $M = N^2$ (N = 1,2,3,…).

The active pumping mechanism could be by condensation or cryosorption at cryogenic temperature for a cryogenic pump. This type of cryogenic pumps has been used in nuclear fusion tokamaks.[7–9,11–13] Another option is the NEG (nonevaporable getter) material, which is widely used as the surface coating material in large vacuum systems, such as the synchrotron,[14–18] or to build a pump by cartridges that integrate many NEG disks on stacks.[19–21] As demonstrated in this study, if the surface coating is after making holes of diameter 0.5 mm and length of 2 mm perpendicular to the surface, which can be easily realized with different mechanical processes or by lasers nowadays, the pumping probability could have a huge increase from initial α = 0.01, 0.02, 0.03 to w = 0.11, 0.17, 0.22, respectively. Compared to the NEG cartridge, the self-replication structure proposed here can mitigate the disadvantage of the shadowing effect between NEG disks in the same cartridge. Because most applications of the gas absorption by NEG or of the cryogenic pumps are for the ultrahigh vacuum system, where the interaction between gas molecules is negligible, the simulation in free molecular flow regime by the TPMC approach in this study is applicable.

Here, we have only demonstrated the impact of this simple self-replication structure on the pumping probability. However, it has great application potentials related to many surface dominated processes, such as isotope separation by membrane,[22,23] hydrogen storage with high surface area,[24,25] heterogeneous catalysis,[26] etc.

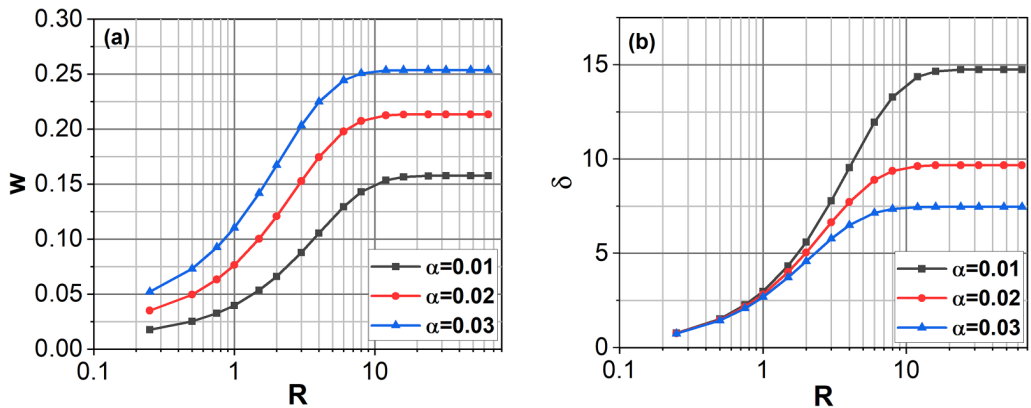

FIG. 3. System pumping probability (a) and its relative increase (b) vs R for different sticking coefficients α and diffuse reflection.

## III. LIGHT ABSORPTION PROBABILITY

Topological effects in photonics have been studied in recent years.[27] The same simulation model in Sec. II will be used to study light absorption and reflection. Only α denotes the absorption rate instead of the sticking coefficient, and the diffuse reflection from active surface is replaced by the specular reflection for photons represented by the test particles in the simulation. Table II lists the simulation results of the absorption probability w of the system and corresponding relative increase $\delta = (w − α)/α$ of w.

The notation of w in Table II is same as in Table I, and the uncertainties in the parentheses are estimated by Eq. (7). From the simulation results, we know that many conclusions in Sec. II hold. However, quantitatively, the specular reflection boundary condition has a greater effect on the improvement of w as shown in Fig. 4.

Compared to the diffuse reflection, we can see that the improvement of w will approach the maximum much slower when the boundary condition is changed to the specular reflection. In our simulation example, the ultimate absorption probability w of the system is given by Eq. (4) by assuming $w_{tube}(α) = 1$, which is w = 0.7875, 0.7897, 0.7918 for α = 0.01, 0.02, 0.03, respectively. It can be seen from the simulation results that for the specular reflection, the maximum values of w can approach the ultimate values as $A/A_s$ or R increases. This implies that $w_{tube}(α)$ will approach unity (blackbody) and is only limited by the normal incident photons from the light source.

In practice, there is no difficulty to make $4 × 10^6$ holes of a diameter of 0.1 mm and a length of 3.2 mm in the original square of a = 200 mm, and the absorption probability w of the system could have a huge increase from initial α = 0.01, 0.02, 0.03 to w = 0.44, 0.59, 0.66, and corresponding relative increase of w is 43 times, 28 times, and 21 times, respectively. The diameter of 0.1 mm is still much larger than the wavelength of light λ, if λ is 600 nm. Under this condition, to calculate the absorption probability by the test particle Monte Carlo simulation without taking the wave diffraction into account would be applicable.[28–30] Obviously, the significant improvement of the system absorption probability by such a simple self-replication geometric structure has great application potential in solar cells if it could be compatible with the need to collect the electrical charges, or in converting the solar irradiation into heat through the collector fluid. Moreover, because the reflection probability of the system is 1 − w, this self-replication geometric structure would be useful to reduce the detection possibility of stealth aircrafts and boats.

In the simulation, the photons (test particles) are coming from a uniform diffuse light source. If the light is the normal incidence, the absorption probability remains as α; if it comes from a special incident angle, the result could be different and is possible to be studied with our simulation model in near future.

## IV. MORE COMPLICATED SYSTEM

As said, if all tubes in each self-replication step N are independent and identical, the effective pumping probability or the absorption probability of the system can be calculated by Eq. (4). Using this method, only the simulation of $w_{tube}(α)$ is needed.

Furthermore, the system in the study is created principally by extruding the circles enclosed by the original squares into a bundle of parallel tubes in the third perpendicular direction. Because only the performance of the tubes has changed, it would be reasonable to reduce the remaining flat area with unchanged performance as much as possible. One possible method to do that is to use tubes of the square cross section instead of tubes of the circular cross section.

Figure 5 shows the simulation results and comparisons of the pumping probability and the absorption probability of these two types of tubes, under diffuse and specular reflections, respectively. Here, R represents the ratio of length to diameter for the circular tube, and the ratio of tube length to square side length for the square tube, respectively. It can also be clearly seen that the final approached maximum value of w depends on the original α, and the greater the original α is, the greater the maximum value of w is. Moreover, the maximum values of w are less than unity under diffuse reflection conditions, but w will approach unity under the specular reflection condition. In addition, compared to the circular tube under diffuse reflection, if the tube is long enough, the maximum values of the pumping probability of the square tube are 4.6%–4.1% higher for α ranging from 0.01 to 0.03. When the diffuse reflection is replaced by the specular reflection, the absorption probability of the square tube is also higher than that of the circular one by up to 3%.

Up to now, the ideal case of no distance between the walls of the adjacent tubes is considered. In this case, the space between the tubes could be used for the installation of necessary heating elements for NEG regeneration or electrodes to collect the electricity or reserved to hold the collector fluid in converting the solar irradiation energy into heat. Actually, it would be more realistic to have a distance between the walls of the adjacent tubes, and Eq. (4) should be modified by using different weighting factors $k_1$ and $k_2$ for $w_{tube}(α)$ and α instead of π/4 and (1 − π/4). For example, if making $1 × 10^6$ circular holes of a diameter of 0.1 mm and a length of 3.2 mm in the original square of a = 200 mm and the distance between adjacent holes about 0.1 mm, and the weighing factor of the opening area for the light is $k_1$ = 0.196 35. In such a case, the photon absorption probability w of the system under specular reflection could have an increase from initial α = 0.01, 0.02, 0.03 to w = 0.117, 0.162, 0.188. If the circular tubes are replaced by the square tubes and the square side length is also 0.1 mm, the corresponding photon absorption probability w of the system will increase to w = 0.150, 0.204, 0.234 from α = 0.01, 0.02, 0.03, respectively. This further increase is mainly because of the larger weighing factor of the opening area for the light $k_1$ = 0.25, which is 27% bigger compared to that for the circular tubes.

In order to exploit the topological effect in the design of an NEG pump and a cryogenic pump, the relationship between the maximum value of $w_{tube}$ of a circular tube and the sticking coefficient α under diffuse reflection and corresponding length to diameter ratio R when $w_{tube}$ reaches 95% of its maximum value are given in Figs. 6(a) and 6(b), respectively.

It can be seen that for large α, the maximum value of $w_{tube}$ will approach unity when the tube is long enough. At a given diameter, the length of the tube needed for $w_{tube}$ to reach 95% of its maximum value becomes shorter as α increases. In addition, as marked in the figure, the nominal sticking coefficient α of the NEG material usually ranges from 0.01 to 0.03,[17] and that of the charcoal coated surface at cryogenic temperature is above 0.5 depending on the gas species.[3] It is seen from Fig. 6 that the benefit, i.e., the relative increase of the pumping probability of the topological structures, is bigger for the NEG material. From the datasheets of commercial NEG pumps and our experience,[20,21] the pumping probability of the NEG pump based on the cartridge structure is less than 0.25 for hydrogen. However, the topological structures provide us an alternative NEG pump design approach, and it would be possible to increase the pumping probability to 0.3 by the square NEG tubes with small distances as shown in Fig. 5(a). More important, the topological structures will greatly improve the pumping ability of NEG coated surfaces.

TABLE II. Absorption probability w and corresponding relative increase δ by specular refection.

| | A | α = 0.01 | δ | α = 0.02 | δ | α = 0.03 | δ |
|---|---|---|---|---|---|---|---|
| Reference case without tube | — | 0.009 999 5(1) | — | 0.019 997 8(1) | — | 0.029 995 6(2) | — |
| N = 1, M = 1 | | | | | | | |
| $T_L = 0.25a$, R = 0.25 | $0.25\pi a^2$ | 0.017 606 4(1) | 0.76 | 0.034 827 2(2) | 0.74 | 0.051 723 4(2) | 0.72 |
| $T_L = 0.5a$, R = 0.5 | $0.5\pi a^2$ | 0.024 930 1(2) | 1.49 | 0.048 735 7(2) | 1.44 | 0.071 651 7(3) | 1.39 |
| $T_L = 0.75a$, R = 0.75 | $0.75\pi a^2$ | 0.032 023 0(2) | 2.20 | 0.061 922 7(2) | 2.10 | 0.090 208 2(3) | 2.01 |
| $T_L = a$, R = 1 | $\pi a^2$ | 0.038 915 1(2) | 2.89 | 0.074 502 5(3) | 2.73 | 0.107 638 9(3) | 2.59 |
| N = 2, M = 4 | | | — | | — | | — |
| $T_L = 0.25a$, R = 0.5 | $0.5\pi a^2$ | 0.024 928 3(2) | 1.49 | 0.048 730 2(2) | 1.44 | 0.071 641 7(3) | 1.39 |
| $T_L = 0.5a$, R = 1 | $\pi a^2$ | 0.038 910 6(2) | 2.89 | 0.074 488 2(3) | 2.72 | 0.107 616 5(3) | 2.59 |
| $T_L = 0.75a$, R = 1.5 | $1.5\pi a^2$ | 0.052 168 8(2) | 4.22 | 0.098 106 6(3) | 3.91 | 0.139 694 9(3) | 3.66 |
| $T_L = a$, R = 2 | $2\pi a^2$ | 0.064 818 9(2) | 5.48 | 0.120 002 7(3) | 5.00 | 0.168 740 7(4) | 4.62 |
| N = 3, M = 16 | | | — | | — | | — |
| $T_L = 0.25a$, R = 1 | $\pi a^2$ | 0.038 904 7(2) | 2.89 | 0.074 472 1(3) | 2.72 | 0.107 588 3(3) | 2.59 |
| $T_L = 0.5a$, R = 2 | $2\pi a^2$ | 0.064 804 3(2) | 5.48 | 0.119 966 8(3) | 5.00 | 0.168 683 3(4) | 4.62 |
| $T_L = 0.75a$, R = 3 | $3\pi a^2$ | 0.088 557 2(3) | 7.86 | 0.159 582 6(4) | 6.98 | 0.219 677 5(4) | 6.32 |
| $T_L = a$, R = 4 | $4\pi a^2$ | 0.110 583 2(3) | 10.06 | 0.194 746 2(4) | 8.74 | 0.263 372 7(4) | 7.78 |
| N = 4, M = 64 | | | — | | — | | — |
| $T_L = 0.25a$, R = 2 | $2\pi a^2$ | 0.064 787 0(2) | 5.48 | 0.119 923 6(3) | 5.00 | 0.168 613 2(4) | 4.62 |
| $T_L = 0.5a$, R = 4 | $4\pi a^2$ | 0.110 543 0(3) | 10.05 | 0.194 657 1(4) | 8.73 | 0.263 243 9(4) | 7.77 |
| $T_L = 0.75a$, R = 6 | $6\pi a^2$ | 0.150 400 9(4) | 14.04 | 0.254 853 4(4) | 11.74 | 0.334 832 6(5) | 10.16 |
| $T_L = a$, R = 8 | $8\pi a^2$ | 0.185 791 9(4) | 17.58 | 0.304 944 9(5) | 14.25 | 0.391 436 3(5) | 12.05 |
| N = 5, M = 256 | | | — | | — | | — |
| $T_L = 0.25a$, R = 4 | $4\pi a^2$ | 0.110 499 3(3) | 10.05 | 0.194 555 3(4) | 8.73 | 0.263 090 7(4) | 7.77 |
| $T_L = 0.5a$, R = 8 | $8\pi a^2$ | 0.185 693 5(4) | 17.57 | 0.304 755 9(5) | 14.24 | 0.391 185 7(5) | 12.04 |
| $T_L = 0.75a$, R = 12 | $12\pi a^2$ | 0.246 300 0(4) | 23.63 | 0.383 909 5(5) | 18.20 | 0.475 270 0(5) | 14.84 |
| $T_L = a$, R = 16 | $16\pi a^2$ | 0.296 760 7(5) | 28.68 | 0.443 936 4(5) | 21.20 | 0.534 658 9(5) | 16.82 |
| N = 6, M = 1024 | | | — | | — | | — |
| $T_L = 0.25a$, R = 8 | $8\pi a^2$ | 0.185 600 0(4) | 17.56 | 0.304 562 2(5) | 14.23 | 0.390 917 2(5) | 12.03 |
| $T_L = 0.5a$, R = 16 | $16\pi a^2$ | 0.296 568 5(5) | 28.66 | 0.443 618 9(5) | 21.18 | 0.534 274 8(5) | 16.81 |
| $T_L = 0.75a$, R = 24 | $24\pi a^2$ | 0.376 350 1(5) | 36.64 | 0.528 477 6(5) | 25.42 | 0.611 613 5(5) | 19.39 |
| $T_L = a$, R = 32 | $32\pi a^2$ | 0.436 904 4(5) | 42.69 | 0.585 131 6(5) | 28.26 | 0.658 691 0(5) | 20.96 |
| N = 7, M = 4096 | | | — | | — | | — |
| $T_L = 0.25a$, R = 16 | $16\pi a^2$ | 0.296 426 3(5) | 28.64 | 0.443 354 8(5) | 21.17 | 0.533 936 5(5) | 16.80 |
| $T_L = 0.5a$, R = 32 | $32\pi a^2$ | 0.436 643 7(5) | 42.66 | 0.584 750 2(5) | 28.24 | 0.658 263 3(5) | 20.94 |
| $T_L = 0.75a$, R = 48 | $48\pi a^2$ | 0.522 329 7(5) | 51.23 | 0.654 015 5(5) | 31.70 | 0.710 393 0(5) | 22.68 |
| $T_L = a$, R = 64 | $64\pi a^2$ | 0.579 596 6(5) | 56.96 | 0.693 347 5(5) | 33.67 | 0.737 017 7(4) | 23.57 |
| N = 8, M = 16 384 | | | | | | | |
| $T_L = 0.25a$, R = 32 | $32\pi a^2$ | 0.438 414 0(2) | 42.84 | 0.587 084 1 (2) | 28.35 | 0.660 820 3(1) | 21.03 |
| $T_L = 0.5a$, R = 64 | $64\pi a^2$ | 0.581 910 9(2) | 57.19 | 0.695 978 2(1) | 33.80 | 0.739 740 7(1) | 23.66 |
| $T_L = 0.75a$, R = 96 | $96\pi a^2$ | 0.652 272 5(2) | 64.23 | 0.736 653 0(1) | 35.83 | 0.764 618 0(1) | 24.49 |
| $T_L = a$, R = 128 | $128\pi a^2$ | 0.692 254 7(1) | 68.23 | 0.755 960 9(1) | 36.80 | 0.775 328 5(1) | 24.84 |
| N = 9, M = 65 536 | | | | | | | |
| $T_L = 0.25a$, R = 64 | $64\pi a^2$ | 0.581 912 9(5) | 57.19 | 0.695 978 9(5) | 33.80 | 0.739 743 3(4) | 23.66 |
| $T_L = 0.5a$, R = 128 | $128\pi a^2$ | 0.692 254 1(5) | 68.23 | 0.755 960 9(1) | 36.80 | 0.775 327 5(4) | 24.84 |
| $T_L = 0.75a$, R = 192 | $192\pi a^2$ | 0.733 558 2(4) | 72.36 | 0.772 862 3(4) | 37.64 | 0.784 004 4(4) | 25.13 |
| $T_L = a$, R = 256 | $256\pi a^2$ | 0.753 190 2(4) | 74.32 | 0.779 731 4(4) | 37.99 | 0.787 311 2(4) | 25.24 |

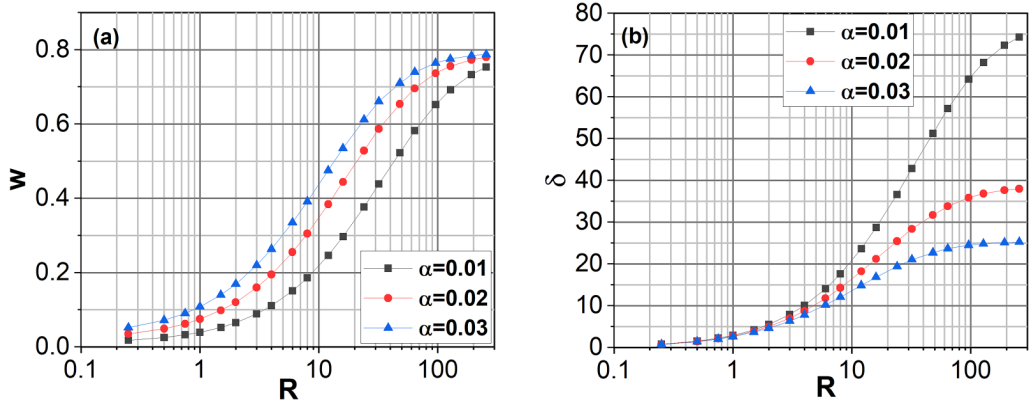

FIG. 4. System absorption probability (a) and its relative increase (b) vs R for different absorption rates $\alpha$ and specular reflection.

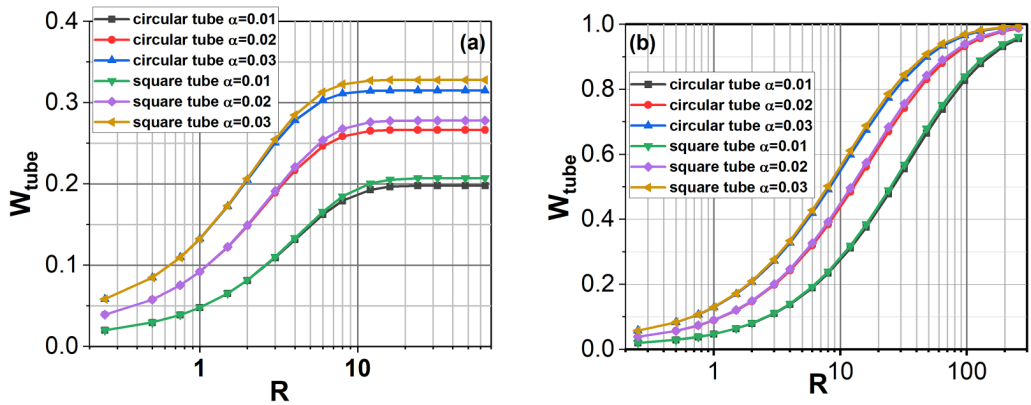

FIG. 5. Pumping probability under diffuse reflection (a) and absorption probability under specular reflection (b) of two different tubes vs R for different sticking coefficients (absorption rates) $\alpha$.

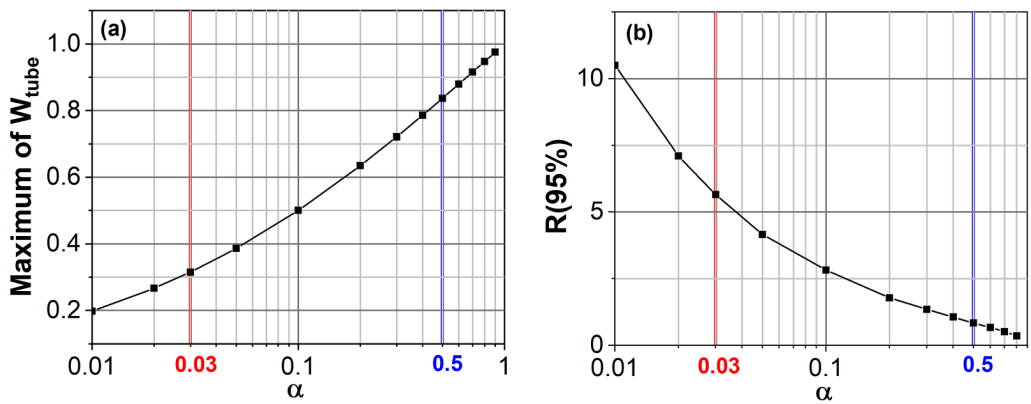

FIG. 6. Maximum of $w_{tube}$ under diffuse reflection (a) and the corresponding R when $w_{tube}$ reaches 95% of the maximum (b) of a circular tube vs the sticking coefficient $\alpha$.

If the system is more complicated and the tubes are not independent of each other, it could be directly simulated. As shown in Table II, the system of 65 536 tubes has been directly and successfully simulated and, in principle, each tube could have individual physical and geometric parameters. In addition, because the code PROVAC3D has a speed-up efficiency of more than 80% with 16 000 cores for parallelization, huge numbers of test particles can be simulated, and the results in this paper are of high statistical precision.

## V. SUMMARY AND CONCLUSIONS

Usually, the requirement of the vacuum pumping is to provide high pumping speed for a given gas load under the system geometric constraint. It is found that under the area constraint of the original square, the topological features of a simple two-dimensional self-replication structure can be exploited by adding tubes in the third perpendicular dimension and the pumping probability of the system can be significantly improved. When the diffuse reflection on surfaces for gas molecules is replaced by the specular reflection for photons, the improvement of the light absorption probability of the system is even greater. Furthermore, the relationship to the geometric and physical parameters and corresponding maximum values of the system pumping/light absorption probability has been obtained by systematic Monte Carlo simulations.

The study in this paper proposes a novel and simple way to achieve better system performance of the pumping probability or the light absorption probability under the geometric constraint and with relatively poor surface pumping sticking coefficient or photon absorption rate. So, the results have a great potential for different applications such as in vacuum pumping, in photovoltaic industry, and by converting solar irradiation energy into heat.

Analog to the cases that originated from a two-dimensional self-replication structure studied in this paper, a three-dimensional structure produced by the self-replication process of a cube with an enclosed sphere has good features that the sum of spheres' surface area can increase as the sum of their volumes remains as the same. One potential application to batteries could increase energy storage density, which is essential for electric cars, assuming that battery storage capabilities are related to the surface area available inside the battery.

## ACKNOWLEDGMENT

We acknowledge EUROfusion for the computation resources allocated to project VAC_ND to run the high performance computer Marconi Fusion at Cineca, Italy.

## DATA AVAILABILITY

The data that support the findings of this study are available from the corresponding author upon reasonable request.